# A highly sensitive magnetic sensor using a 2D van der Waals ferromagnetic material

Valery Ortiz Jimenez[1], Vijaysankar Kalappattil[1], Tatiana Eggers[1], Manuel Bonilla[1], Sadhu Kolekar[1], Pham Thanh Huy[2], Matthias Batzill[1], and Manh-Huong Phan[1,a]

[1] Department of Physics, University of South Florida, Tampa, FL 33620, USA

[2] Phenikaa Institute for Advanced Study, Phenikaa University, Yen Nghia, Ha-Dong District, Hanoi 1000, Viet Nam

Two-dimensional (2D) van der Waals ferromagnetic materials are emerging as promising candidates for applications in ultra-compact spintronic nanodevices, nanosensors, and information storage. Our recent discovery of the strong room temperature ferromagnetism in single layers of $VSe_2$ grown on graphite or $MoS_2$ substrate has opened new opportunities to explore these ultrathin magnets for such applications. In this paper, we present a new type of magnetic sensor that ultilizes the single layer $VSe_2$ film as a highly senstive magnetic core. The sensor relies in changes in resonance frequency of the LC circuit composed of a soft ferromagnetic microwire coil that contains the ferromagnetic $VSe_2$ film subject to applied DC magnetic fields. The senstivity of the sensor reaches an extremely high value of $16 \times 10^6$ Hz/Oe, making it an excellent candidate for a wide range of magnetic sensing applications.

**PACS:** 75.50.Kj

**Keywords:** Monolayer magnets; Magneto-LC resonance; Magnetic sensors

Since intrinsic long-range ferromagnetic order was realized in 2017 with bulk exfoliated monolayers of $Cr_2Ge_2Te_6$ and $CrI_3$,[1,2] the potential of two-dimensional (2D) van der Waals magnets has excited the scientific community.[3-11] While these 2D magnets have demonstrated their usefulness in magnetoelectric devices, their technological applications are restricted to low temperatures (< 100 K).[6-10] In contrast, recent discoveries of the room temperature (RT) ferromagnetism in transition metal dichalcogenide (TMD) monolayers of $VSe_2$ and $MnSe_2$ grown by van der Waals epitaxy on various substrates (graphite, $MoS_2$, GaSe) may enable applications at ambient temperature.[3,4,10]

In this paper, we present a new type of magnetic sensor that intergrates a single layer $VSe_2$ film within a Co-rich microwire-based coil. We note that induction coil sensors have been widely used due to the simplicity of their construction and a well known transfer function.[12] It is established that the sensivity of an induction coil is limited by the number of windings; the greater the number of turns the higher the sensitivity. This quickly becomes a problem for modern applications where it is desirable to limit the size of sensors. Adding a soft ferromangetic core with high relative permeability can largely increase the sensitivity of a coil, and hence allow for smaller sensor sizes.[13] A similar working principle has recently been used in the design of MMC-LC resonator sensors, but cobalt rich magnetic microwires are used instead of non-magnetic conductors such as copper.[14] The working principle of the sensor reported in this paper is fundamentally different than that of a conventional induction coil sensor. It relies on the changes in resonance frequency caused by external magnetic fields, instead of simply measuring the induction of the coil. It is also different from the MMC-LC resonator sensor[14] that relies in changes in impedance of the microwire caused by external magnetic fields.

A simple model for a coil sensor is a lumped element representation of a non-ideal inductor. Winding cylindrical conductors close to each other will introduce parasitic elements $R_{par}$ and $C_{par}$, such that the non-ideal inductor can be represented as a series combination of an ideal inductor $L$ and $R_{par}$, in parallel with $C_{par}$.[15,16,17] The impedance of the coil $Z_{coil}$ can then be written as

$$Z_{\text{coil}} = (Z_{R_{\text{par}}} + Z_L) \,||\, Z_{C_{\text{par}}} \tag{1}$$

$$Z_{\text{coil}} = \frac{1}{\frac{1}{R_{\text{par}}+j\omega L}+\frac{1}{1/j\omega C_{\text{par}}}} \tag{2}$$

$$Z_{\text{coil}} = \frac{R_{par}+j\omega[L(1-\omega^2 L C_{par})-C_{par}R_{par}^2]}{(1-\omega^2 L C_{par})^2+(\omega C_{par}R_{par})^2}, \tag{3}$$

where $\omega$ is the angular frequency, and $j$ is the imaginary unit. Resonance will occur when the inductive reactance ($X_L$) and the capacitive reactance ($X_C$) are equal in magnitude but differ in phase by 180 degrees. At this point very little current flows through the wire, the impedance $Z_{Coil}$ becomes very large and self-resonance is achieved.[18,19] The resonance frequency is given by

$$f_0 = \frac{\sqrt{1-(R_{\text{par}}^2 C_{\text{par}}/L)}}{2\pi\sqrt{LC_{\text{par}}}}, \tag{4}$$

where $f_0$ is the resonance frequency. We expect a self-resonant behavior in the MMC as well, but the resonance frequency will be different than that of the inductor since the wire is now a magnetic material. We must also consider the effects of a ferromagnetic core on the sensor. The core will modify the relative permeability in the space within the coil, and this will in turn change the flux through the coil and hence affect the inductance. Since the permeability is field dependent, an external magnetic field will modify the permeability of the core and the resonance frequency with it. The sensitivity of the sensor is defined as the rate of change of the resonance frequency with respect to the external DC magnetic field,

$$Sensitivity = \frac{df_0}{dH}. \qquad (5)$$

The *Q* factor is also calculated by measuring the bandwidth, BW, and the resonance frequency using the following relation,

$$Q = \frac{f_0}{BW} \qquad (6)$$

To make the magnetic sensor, a $Co_{69.25}Fe_{4.25}Si_{13}B_{12.5}Nb_1$ wire (diameter, ~60 μm) was wound into a 15-turn, 10 mm long coil with a 5 mm internal diameter. The fabrication details and material characterization of the microwire can be found elsewhere.[20,21] Monolayer $VSe_2$ was used as the core of the coil. The single layer films of $VSe_2$ were grown on graphite (HOPG) and single crystal $MoS_2$ by molecular beam epitaxy (MBE); the details of which have been reported in our previous work.[3] Since both monolayer $VSe_2$ samples grown on $MoS_2$ and HOPG show similar magnetic and sensing properties, in this paper we only report on the properties of monolayer $VSe_2$ grown on $MoS_2$. Figure 1(a) shows a typical scanning tunnel microscopy (STM) image of the $VSe_2$ film. It can be seen that the single layer of $VSe_2$ was epitaxially grown on the $MoS_2$ substrate. The magnetization versus magnetic field (*M-H*) curves, measured by a vibrating sample magnetometer (VSM) on $VSe_2$ grown on both HOPG and $MoS_2$, show a soft ferromagnetic characteristic (small coercive field, small remanent magnetization, and high saturation magnetization) of the single layer $VSe_2$ film at room temperature (Fig. 1(b)), which is desirable for its use as the core of the sensor that itself operates at room temperature. The core was ~14 mm long and ~0.4 mm wide, allowing us to insert and remove the core with ease. The sensor was mounted on a test fixture made of a dielectric material on top of a copper ground plane; SMA connector ports were soldered to the ground plane on two ends and the two leads of the inductor were soldered to the center pin of each connector.

One port reflection measurements were performed with an HP 4191A impedance analyzer over the frequency range 1 – 200 MHz. A simple short-open-load (SOL) calibration was made. The test fixture was connected to the impedance analyzer through a coaxial cable, and it was terminated with a 50 ohm cap. The effects of the cable were removed by the calibration. The impedance (*Z*), resistance (*R*), and reactance (*X*) are calculated from the reflection measurement. An external DC magnetic field was generated by a Helmholtz coil. A diagram of the setup is shown in Fig. 1(c,d). The external field was applied transverse to the coil axis as the frequency was swept. A frequency sweep was performed for every value of the field which was swept from -30 to 30 Oe in steps of 1 Oe.

A set of measurements is performed to characterize the sensor; the resonance frequency is found for every value of the field. Resonance is determined by the zero crossing of the reactance. To see the effects of the monolayer core we show in Fig. 2(a) the reactance of the coil with and without the monolayer core at zero field. The presence of the monolayer shifts the resonance frequency of the sensor by a few megahertz. We then applied the external field and observed how the reactance curve is shifted along with the corresponding resonance frequency as shown in Fig. 2(b), for the coil with the monolayer in its core. The reactance of the coil being proportional to the inductance changes drastically with frequency and applied field, as well as the resonance frequency (see Fig. (2c)).

The resonance frequency is determined for every value of the field. Figure 3(a) shows the change in resonance frequency as a function of applied field. Unique values of the resonance frequency exist for fields smaller than 5 Oe. At higher fields, the magnetic wire saturates, and $f_0$ changes very slowly, rending the sensor inoperable. A fit of the right branch of the $f_0$ plot is shown in Fig. 3(b) as well as the sensitivity of the sensor. Values as large as $16\times10^6$ Hz/Oe are achieved for the sensitivity. The massive value of the sensitivity shows that the present sensor can be used to accurately determine the magnitude of small magnetic fields as well as subtle variations in them. This brings forth a wide range of applications such as bio-detection, where magnetic nanoparticles are used as bio-markers for pathogen detection and quantifying low particle concentrations is neccessay.[22] Finally, we have determined the Q factor of the sensor at different values of external DC magnetic field, as shown in Fig. 4. We have found that for very small fields, smaller than 1 Oe, the Q factor is 0.6; as the field increases it reaches a minimum value of 0.36 at 3 Oe. Most of the losses are associated with the magnetic wire itself; the core slightly decreases the Q factor by less than 0.1. This is consistent with the soft ferromagnetic characteristic of monolayer $VSe_2$ (Fig. 1(b)), which predicts minimal magnetic losses.

In summary, a highly sensitive magnetic field sensor was built using a magnetic microwire coil as the sensing element with a single layer $VSe_2$ film as the core. Sensitivity as large as $16\times10^6$ Hz/Oe for very small fields was obtained. We successfully implemented $VSe_2$ monolayers in room temperature magnetic field sensors. Furthermore, core losses were shown to be minimal which is desirable for sensor design.

**ACKNOWLEDGMENTS**

Work was supported by the VISCOSTONE USA under Grant No. 1253113200 and the National Science Foundation under grants DMR-1701390 and ECCS-1608654. T.E. also acknowledges support from NASA Florida Space Grant Consortium (FSGC) under Award No. 1253-1124-00.

**Figure captions**

**FIG. 1.** (a) A typical STM image of the single layer VSe$_2$ film grown on single crystal MoS$_2$; (b) the magnetic hysteresis (*M-H*) loop of the film measured at 300 K; (c) A block diagram of the measurement setup for testing performances of the sensor; (d) the LC-circuit composed of a magnetic microwire with the VSe$_2$ film inserted in it.

**FIG. 2.** (a) Frequency dependence of the reactance of the sensor with and without the magnetic VSe$_2$ film in the absence of an external magnetic field; (b) Frequency dependence of the reactance of the sensor with the magnetic VSe$_2$ film shows large shifts in the resonant frequency with respect to applied magnetic fields; (c) 2D surface plot shows the magnetic field and frequency dependences of reactance.

**FIG. 3.** (a) The resonant frequency of the sensor ($f_0$) changes as a function of magnetic field. The extremely large change is observed at low fields; (b) A fit of the magnetic field dependent resonant frequency $f_0$(H) data is displayed, with the inset showing the magnetic field dependence of the sensor sensitivity. The extremely large value of the sensor sensitivity indicates the usefulness of the sensor for a wide range of sensing applications.

**FIG. 4.** Magnetic field dependence of the Q factor of the sensor. The significant change in Q is observed at low field regime.

**Figure 1**

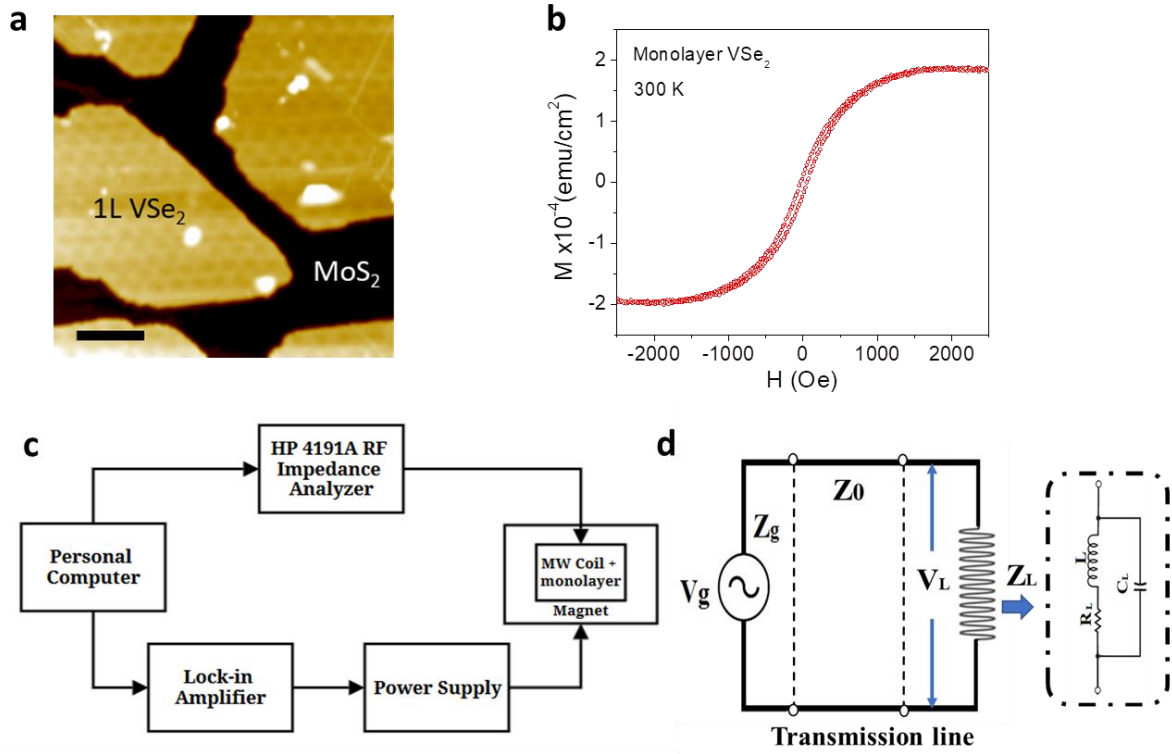

**Figure 2**

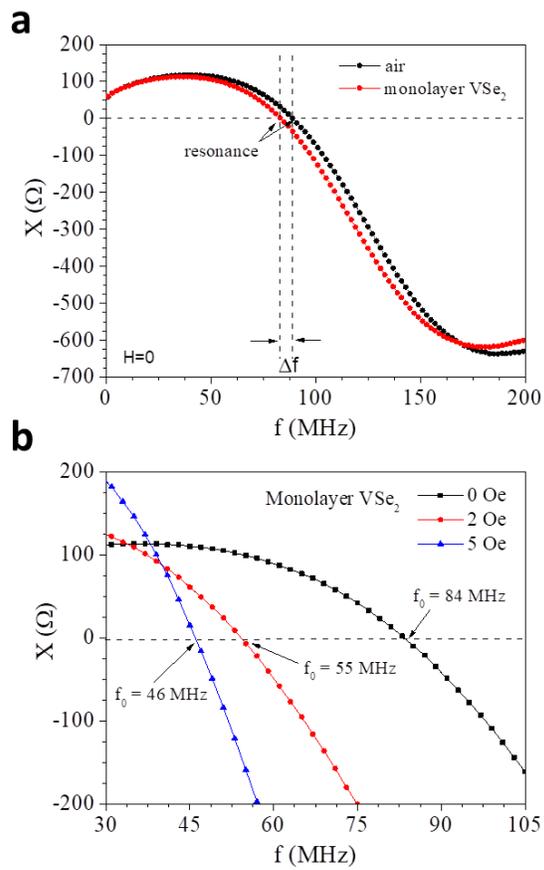
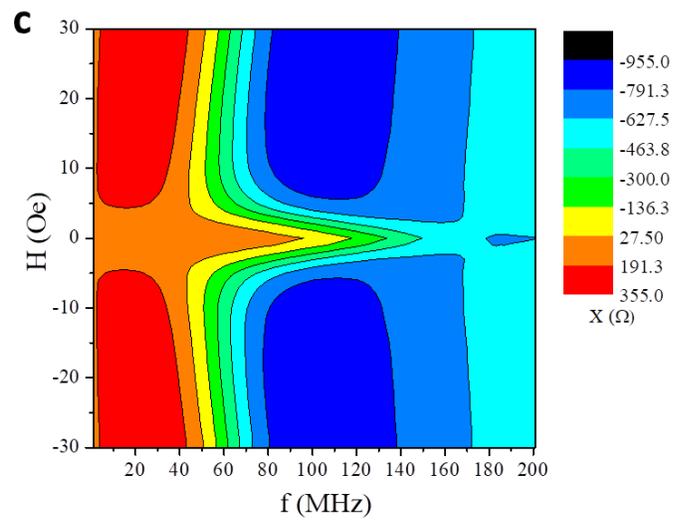

**Figure 3**

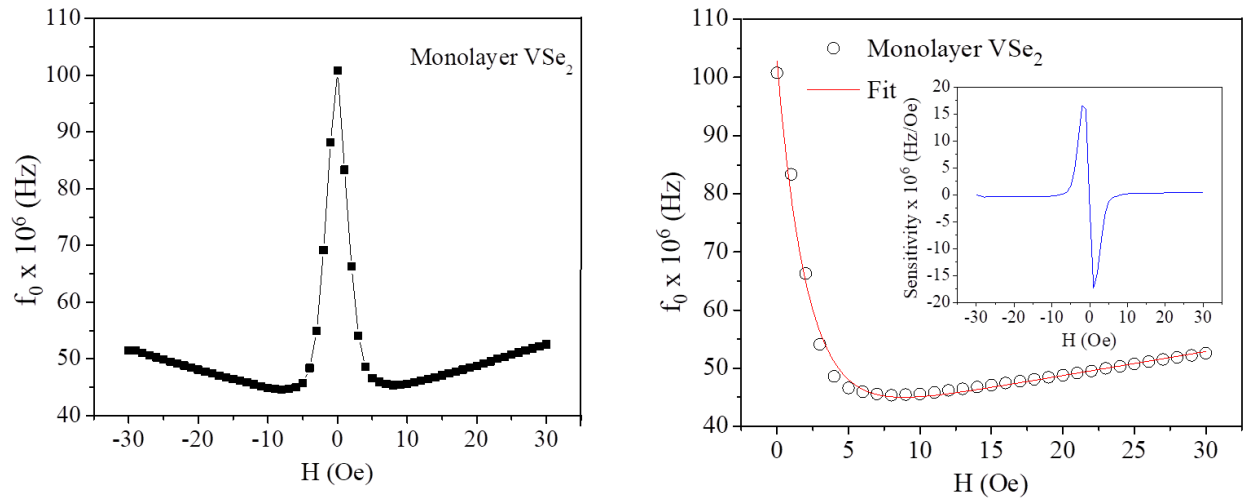

**Figure 4**

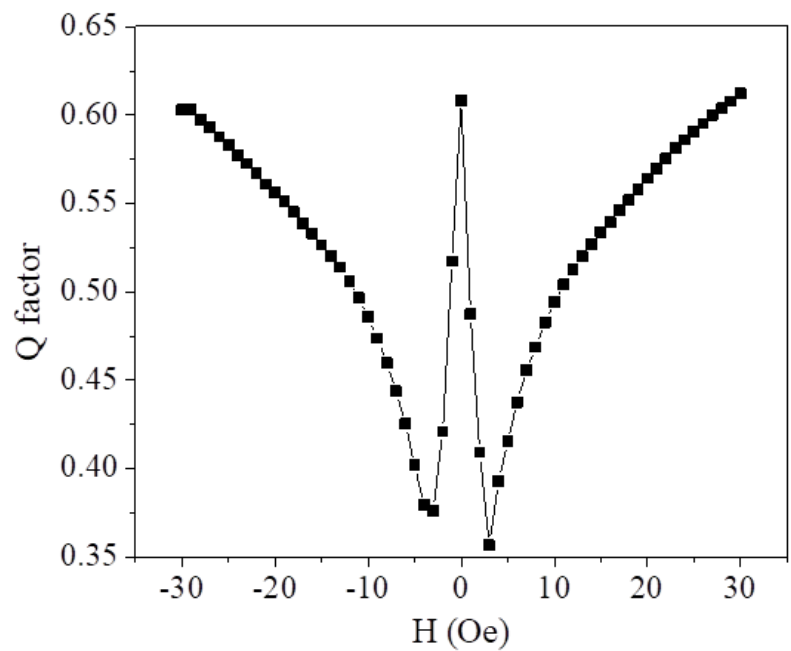